\documentclass{article} % use larger type; default would be 10pt
\usepackage[utf8]{inputenc} % set input encoding (not needed with XeLaTeX)
\usepackage{amsmath}
\usepackage{amsthm} 
\usepackage{amssymb} 
\usepackage{ setspace}
\usepackage{graphicx}
\usepackage[T1]{fontenc}

\usepackage[bitstream-charter]{mathdesign}

\usepackage[english]{babel}
\usepackage{supertabular}
\usepackage{multicol}
\usepackage{breqn}
\usepackage{xcolor}

\newcommand{\mathsym}[1]{{}}
\newcommand{\unicode}[1]{{}}

\usepackage[top=1.8in, bottom=1.8in, left=1.2in, right=1.2in]{geometry}

\usepackage{booktabs} % for much better looking tables
\usepackage{array} % for better arrays (eg matrices) in maths
\usepackage{paralist} % very flexible & customisable lists (eg. enumerate/itemize, etc.)
\usepackage{verbatim} % adds environment for commenting out blocks of text & for better verbatim
\usepackage{subfig} % make it possible to include more than one captioned figure/table in a single float
\usepackage{fancyhdr} % This should be set AFTER setting up the page geometry
\usepackage{sectsty}
\usepackage[nottoc,notlof,notlot]{tocbibind} % Put the bibliography in the ToC
\usepackage[titles,subfigure]{tocloft} % Alter the style of the Table of Contents

\sectionfont{\large}

\title{Four Points Beginner Risk Managers Should Learn from Jeff Holman's Mistakes in the Discussion of \textit{Antifragile}} 
\author{Nassim Nicholas Taleb }
\date{January 2014}

\begin{document}
\maketitle
\setstretch{1.1}
\begin{abstract} 
\noindent
Using Jeff Holman's comments in \textit{Quantitative Finance} to illustrate 4 critical errors students should learn to avoid: 1) Mistaking tails (4th moment) for volatility (2nd moment), 2) Missing Jensen's Inequality, 3) Analyzing the hedging wihout the underlying, 4) The necessity of a num\'eraire in finance.
\end{abstract}

\maketitle % Print the title and abstract box

%\tableofcontents % Print the contents section

\thispagestyle{empty} % Removes page numbering from the first page
\noindent
The review of \textit{Antifragile} by Mr Holman (dec 4, 2013) is replete with factual, logical, and analytical errors. We will only list here the critical ones, and ones with generality to the risk management and quantitative finance communities; these should be taught to students in quantitative finance as central mistakes to avoid, so beginner quants and risk managers can learn from these fallacies. 

\section{Conflation of Second and Fourth Moments}
It is critical for beginners not to fall for the following elementary mistake. Mr Holman gets the relation of the VIX (volatility contract) to betting on "tail events" backwards. Let us restate the notion of "tail events" in the \textit{Incerto} (that is the four books on uncertainty of which \textit{Antifragile} is the last installment): it means a disproportionate role of the tails in defining the properties of distribution, which, mathematically, means a smaller one for the "body".\footnote{The point is staring at every user of spreadsheets: Kurtosis, or scaled fourth moment, the standard measure of fattailedness, entails normalizing the fourth moment by the square of the variance.}

Mr Holman seems to get the latter part of the attributes of fattailedness in reverse. It is an error to mistake the VIX for tail events. The VIX is mostly affected by at-the-money options which corresponds to the center of the distribution, closer to the second moment not the fourth (at-the-money options are actually linear in their payoff and correspond to the conditional first moment). As explained about seventeen years ago in \textit{Dynamic Hedging} (Taleb, 1997) (see appendix), in the discussion on such tail bets, or "fourth moment bets", betting on the disproportionate role of tail events of fattailedness is done by \textit{selling} the around-the-money-options (the VIX) and purchasing options in the tails, in order to extract the second moment and achieve neutrality to it (sort of becoming "market neutral"). Such a neutrality requires some type of "short volatility" in the body because higher kurtosis means lower action in the center of the distribution.

 A more mathematical formulation is in the technical version of the \textit{Incerto} : fat tails means "higher peaks" for the distribution as, the fatter the tails, the more markets spend time between $\mu -\sqrt{\frac{1}{2} \left(5-\sqrt{17}\right)} \sigma$ and $\mu +\sqrt{\frac{1}{2} \left(5-\sqrt{17}\right)} \sigma $ where $\sigma$ is the standard deviation and $\mu$ the mean of the distribution (we used the Gaussian here as a base for ease of presentation but the argument applies to all unimodal distributions with "bell-shape" curves, known as semiconcave). And "higher peaks" means less variations that are not tail events, more quiet times, not less. For the consequence on option pricing, the reader might be interested in a quiz I routinely give students after the first lecture on derivatives: "What happens to at-the-money options when one fattens the tails?", the answer being that they should drop in value. \footnote{\textbf{Technical Point: Where Does the Tail Start?} For a general class of symmetric distributions with power laws, the tail starts at: $\pm\frac{\sqrt{\frac{5 \alpha +\sqrt{(\alpha +1) (17 \alpha +1)}+1}{\alpha -1}} s}{\sqrt{2}}$, with $\alpha$ infinite in the stochastic volatility Gaussian case and $s$ the standard deviation. The "tail" is located between around 2 and 3 standard deviations. This flows from the heuristic definition of fragility as second order effect: the part of the distribution is convex to errors in the estimation of the scale. But in practice, because historical measurements of STD will be biased lower because of small sample effects (as we repeat fat tails accentuate small sample effects), the deviations will be $>$ 2-3 STDs.}
%\end{mdframed}
%\end{wrapfigure}

Effectively, but in a deeper argument, in the QF paper (Taleb and Douady 2013), our measure of fragility has an opposite sensitivity to events around the center of the distribution, since, by an argument of survival probability, what is fragile is sensitive to tail shocks and, critically, should not vary in the body (otherwise it would be broken).

\section{Missing Jensen's Inequality in Analyzing Option Returns} 
Here is an error to avoid at all costs in discussions of volatility strategies or, for that matter, anything in finance. Mr Holman seems to miss the existence of Jensen's inequality, which is the entire point of owning an option, a point that has been belabored in \textit{Antifragile}. One manifestation of missing the convexity effect is a critical miscalculation in the way one can naively assume options respond to the VIX. 
\begin{quote}
"A \$1 investment on January 1, 2007 in a strategy of buying and rolling short-term VIX futures would have peaked at \$4.84 on November 20, 2008 -and then subsequently lost 99\% of its value over the next four and a half years, finishing under \$0.05 as of May 31, 2013." \footnote{In the above discussion Mr Holman also shows evidence of dismal returns on index puts which, as we said before, respond to volatility not tail events. These are called, in the lingo, "sucker puts".}
\end{quote}

This mistake in the example given underestimates option returns by up to... several orders of magnitude. Mr Holman analyzes the performance a tail strategy using investments in financial options by using the VIX (or VIX futures) as proxy, which is mathematically erroneous owing to second- order effects, as the link is tenuous (it would be like evaluating investments in ski resorts by analyzing temperature futures).  Assume a periodic rolling of an option strategy: an option 5 STD away from the money \footnote{We are using implied volatility as a benchmark for its STD.} gains 16 times in value if its implied volatility goes up by 4, but only loses its value if volatility goes to 0. For a 10 STD it is 144 times. And, to show the acceleration, assuming these are traded, a 20 STD options by around 210,000 times\footnote{An event this author witnessed, in the liquidation of Victor Niederhoffer, options sold for \$.05 were purchased back at up to \$38, which bankrupted Refco, and, which is remarkable, without the options getting close to the money: it was just a panic rise in implied volatility.}. 
There is a second critical mistake in the discussion: Mr Holman's calculations here exclude the payoff from actual in-the-moneyness.

One should remember that the VIX is not a price, but an inverse function, an index derived from a price: one does not buy "volatility" like one can buy a tomato; operators buy options correponding to such inverse function and there are severe, very severe nonlinearities in the effect. Although more linear than tail options, the VIX is still convex to actual market volatility, somewhere between variance and standard deviation, since a strip of options spanning all strikes should deliver the variance (Gatheral,2006). The reader can go through a simple exercise. Let's say that the VIX is "bought" at 10\% -that is, the component options are purchased at a combination of volatilities that corresponds to a VIX at that level. Assume returns are in squares. Because of nonlinearity, the package could benefit from an episode of 4\% volatility followed by an episode of 15\%, for an average of 9.5\%; Mr Holman believes or wants the reader to believe that this 0.5 percentage point should be treated as a loss when in fact second order un-evenness in volatility changes are more relevant than the first order effect.

\section{The Inseparability of Insurance and Insured}
One should never calculate the cost of insurance without offsetting it with returns generated from packages than one would not have purchased otherwise.

Even had he gotten the sign right on the volatility, Mr Holman in the example above analyzes the performance of a strategy buying options to protect a tail event without adding the performance of the portfolio itself, like counting the cost side of the insurance without the performance of what one is insuring that would not have been bought otherwise. Over the same period he discusses the market rose more than 100\%: a healthy approach would be to compare dollar-for-dollar what an investor would have done (and, of course, getting rid of this "VIX" business and focusing on very small dollars invested in tail options that would allow such an aggressive stance). Many investors (such as this author) would have stayed out of the market, or would not have added funds to the market, without such an insurance. 

\section{The Necessity of a Num\'eraire in Finance}
There is a deeper analytical error. 

A barbell is defined as a bimodal investment strategy, presented as investing a portion of your portfolio in what is explicitly defined as a "num\'eraire repository of value" (\textit{Antifragile}), and the rest in risky securities (\textit{Antifragile} indicates that such num\'eraire would be, among other things, inflation protected). Mr Holman goes on and on in a nihilistic discourse on the absence of such riskless num\'eraire (of the sort that can lead to such sophistry as "he is saying one is safer on \textit{terra firma} than at sea, but what if there is an earthquake?"). 

The familiar Black and Scholes derivation uses a riskless asset as a baseline; but the literature since around 1977 has substituted the notion of "cash" with that of a num\'eraire , along with the notion that one can have different currencies, which technically allows for changes of probability measure. A num\'eraire is defined as \textit{the unit to which all other units relate}. ( Practically, the num\'eraire is a basket the variations of which do not affect the welfare of the investor.) Alas, without num\'eraire, there is no probability measure, and no quantitative in quantitative finance, as one needs a unit to which everything else is brought back to. In this (emotional) discourse, Mr Holton is not just rejecting the barbell per se, but any use of the expectation operator with any economic variable, meaning he should go attack the tens of thousand research papers and the existence of the journal \textit{Quantitative Finance} itself.

Clearly, there is a high density of other mistakes or incoherent statements in the outpour of rage in Mr Holman's review; but I have no doubt these have been detected by the \textit{Quantitative Finance} reader and, as we said, the object of this discussion is the prevention of analytical mistakes in quantitative finance.

To conclude, this author welcomes criticism from the finance community that are not straw man arguments, or, as in the case of Mr Holmam, violate the foundations of the field itself.

\begin{figure}
\includegraphics[width=.7\linewidth]{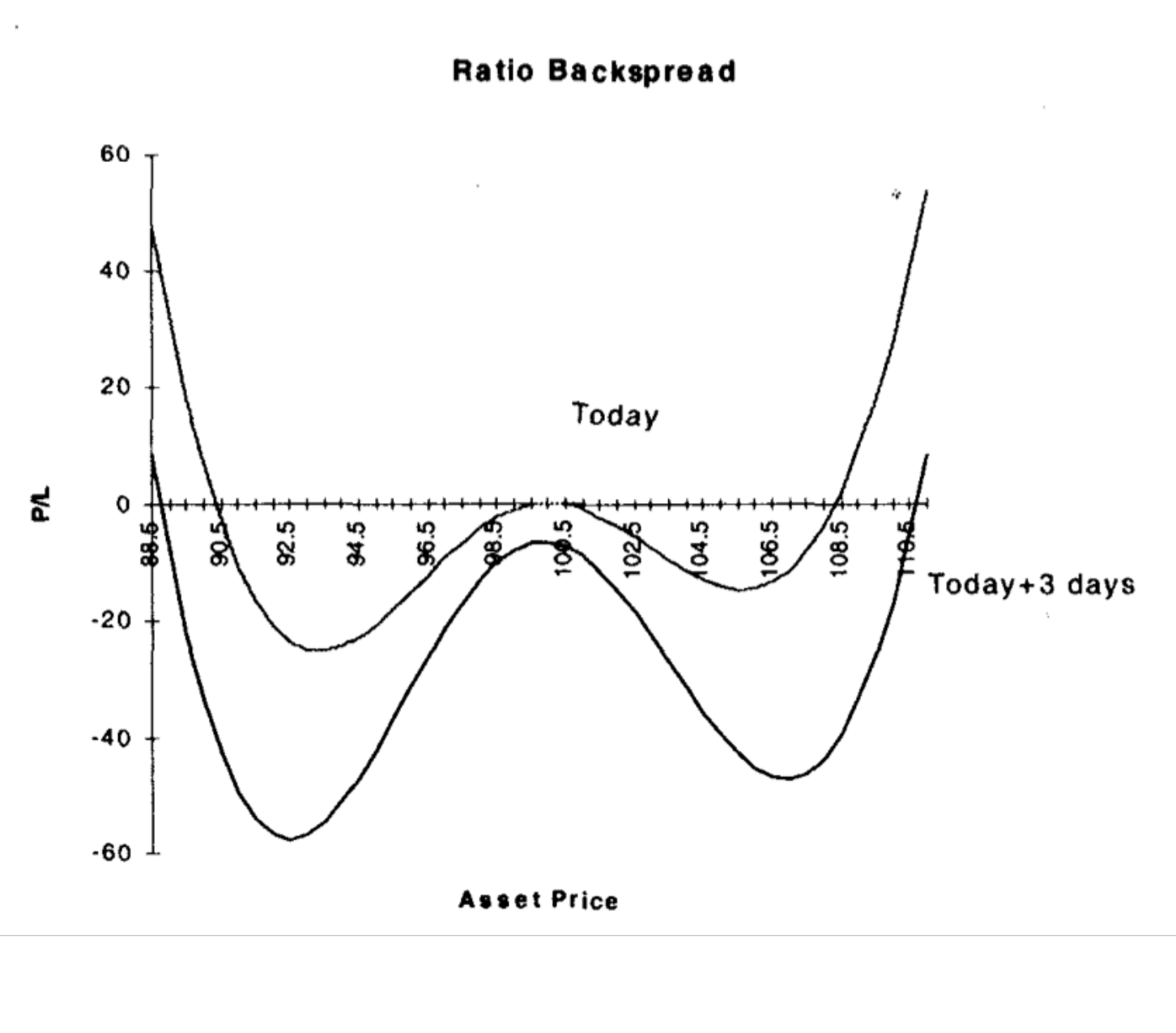}
\caption{First Method to Extract the Fourth Moment, from \textit{Dynamic Hedging, 1997.}}\label{ratiobackspread}
\end{figure}

\begin{figure}
\includegraphics[width=.7\linewidth]{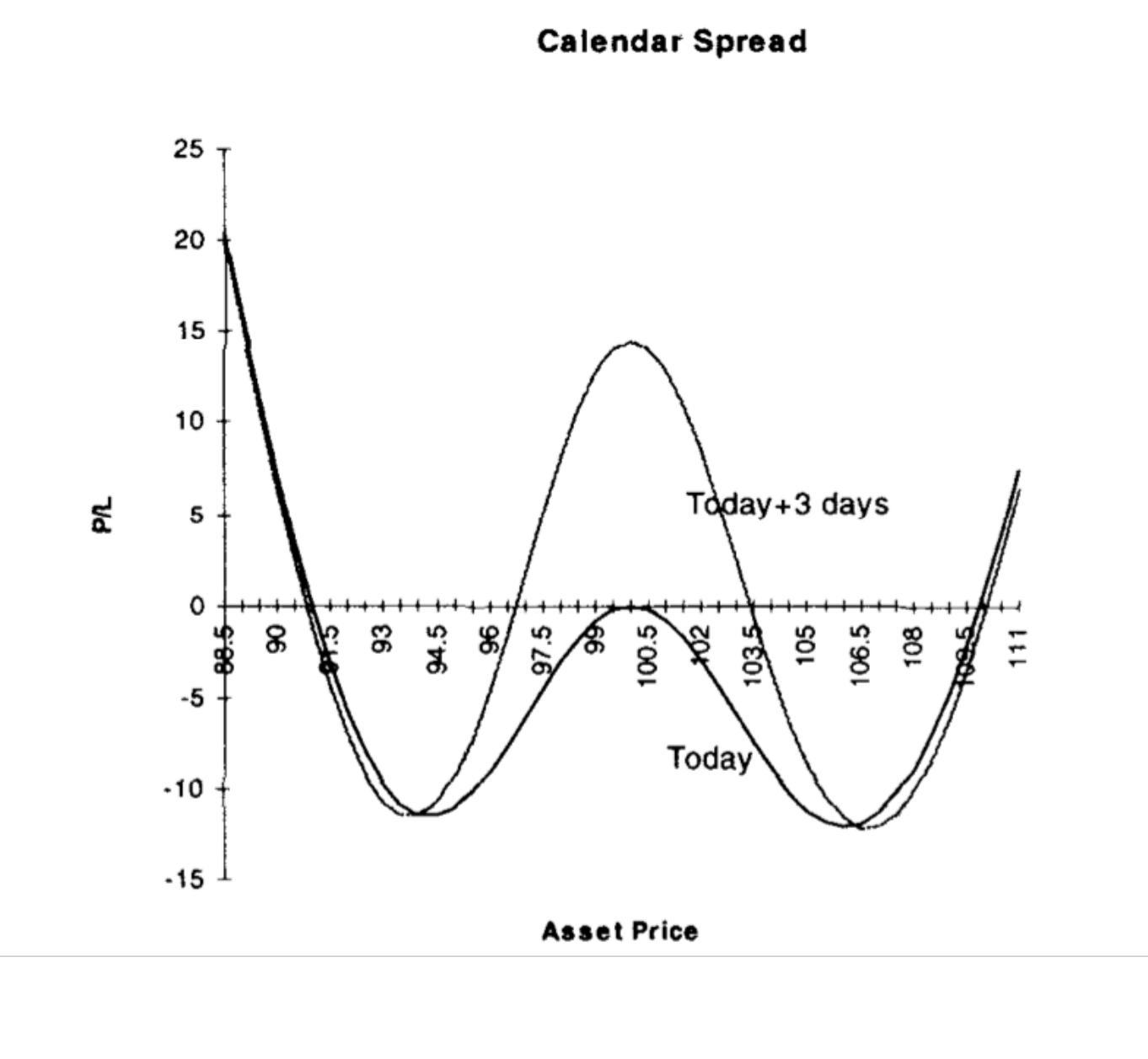}
\caption{Second Method to Extract the Fourth Moment , from \textit{Dynamic Hedging, 1997.}}\label{calendarspread}
\end{figure}

\section*{Appendix (Discussion of Betting on Tails of Distribution in \textit{Dynamic Hedging, 1997})}

From \textit{Dynamic Hedging}, pages 264-265:

\begin{quote}

\textit{A fourth moment bet is long or short the volatility of volatility. It could be achieved either with out-of-the-money options or with calendars. 
Example: A ratio "backspread" or reverse spread is a method that includes the buying of out-of-the-money options in large amounts and the selling of smaller amounts of at-the-money but making sure the trade satisfies the "credit" rule (i.e., the trade initially generates a positive cash flow). The credit rule is more difficult to interpret when one uses in-the-money options. In that case, one should deduct the present value of the intrinsic part of every option using the put-call parity rule to equate them with out-of-the-money.}

\textit{The trade shown in Figure 1 was accomplished with the purchase of both out-of-the-money puts and out-of-the-money calls and the selling of smaller amounts of at-the-money straddles of the same maturity.}

\textit{Figure 2 shows the second method, which entails the buying of 60- day options in some amount and selling 20-day options on 80\% of the amount.
Both trades show the position benefiting from the fat tails and the high peaks. Both trades, however, will have different vega sensitivities, but close to flat modified vega.}
\end{quote}

\section*{Appendix: The Body, The Shoulders, and The Tails}\label{crossoverandtunnel}
From the new book (Taleb 2014): We assume tails start at the level of convexity of the segment of the probability distribution to the scale of the distribution.

\subsection* {Where Do the Tails Start? The Crossovers and Tunnel Effect.}

Notice in Figure 3 a series of crossover zones, invariant to \(a\). Distributions called "bell shape" have a convex-concave-convex
shape (or quasi-concave shape).

\begin{figure*}[ht]
%\begin{center}
\includegraphics[width=\linewidth]{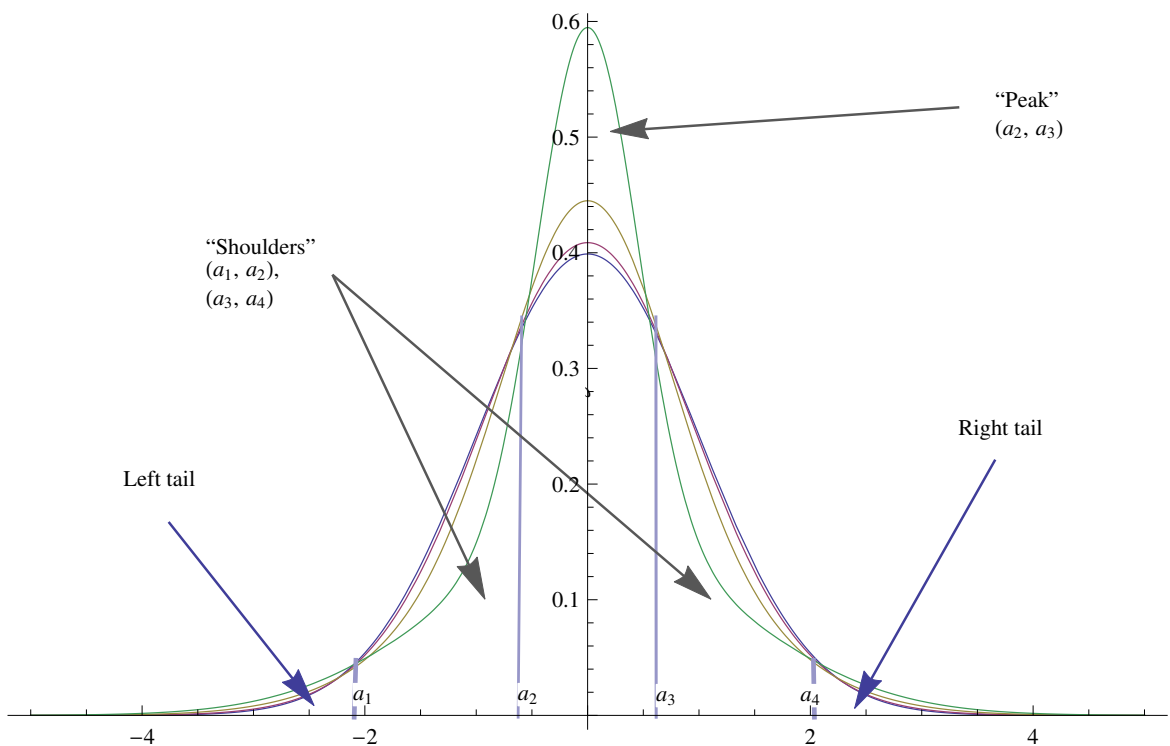}
\caption{Fatter and Fatter Tails through perturbation of $\sigma$. The  mixed distribution with stochastic volatility coefficient  $\{v_i\}_{i=1}^4= \{0,\frac{1}{4},\frac{1}{2},\frac{3}{4}\}$. We can see crossovers $a_1$ through $a_4$. The "tails" proper start at $a_4$ on the right and $a_1$on the left.}\label{crossovers}
%\%end{center}
\end{figure*}

\ \
Let X be a random variable, the distribution of which $p(x)$ is from a general class of all unimodal one-parameter continous pdfs \(p_{\sigma
}\) with support $\mathcal{D}$ $\subseteq $ $\mathbb{R}$ and scale parameter $\sigma $. Let $p(.)$ be quasi-concave on the domain, but neither convex nor concave. The density function $p(x)$ satisfies: \(p(x)\geq p(x+\epsilon)\) for all \(\epsilon>0\), and \(x> x^*\) and \(p(x)\geq p(x-\epsilon)\) for all \(x < x^*\) with \( \{x^*:p(x^*)=\mathrm{max}_x\  p(x)\}\). The class of quasiconcave functions is defined as follows: for all $x$ and $y$ in the domain and  \( \omega \in [0,1]\),
\[p \left( \omega \, x + (1 - \omega) \ y \right) \geq\min \left( p(x),p(y) \right) \]

1- If the variable is "two-tailed", that is, $\mathcal{D}$= (-$\infty $,$\infty $), where \(p^{\delta }( x)\) $\equiv $ \(\frac{p( x+\delta
)+p(x-\delta )}{2}\)
\begin{enumerate}

\item There exist a "high peak" inner tunnel, \(A_T\)= (\(\left.a_2, a_3\right)\) for which the $\delta $-perturbed $\sigma $ of the probability
distribution \(p^{\delta }( x)\)$\geq $\(p (x)\) if \textit{ x} $\in $ (\(\left.a_2, a_3\right)\)

\item There exists outer tunnels, the "tails", for which \(p^{\delta }( x)\)$\geq $\(p (x)\) if \textit{ x} $\in $ ($-\infty $, \(a_1\)) or \(x\)
$\in $ (\(a_4, \infty\))

\item There exist intermediate tunnels, the "shoulders", where \(p^{\delta }( x)\)$\leq $ \(p (x)\) if \textit{ x} $\in $ \(\left(a_1, a_2\right.\))
or \textit{ x} $\in $ \(\left(a_3, a_4\right.\))

A=$\{$\(a_i\)$\}$ is the set of solutions \(\left\{x:\frac{\partial ^2p(x)}{\partial \sigma \, ^2}|_a=0\right\}\). 
\end{enumerate}
For the Gaussian ($\mu $, $\sigma $), the solutions are obtained by setting the second derivative to 0, so
\[\frac{e^{-\frac{(x-\mu )^2}{2 \sigma ^2}} \left(2 \sigma ^4-5 \sigma ^2 (x-\mu )^2+(x-\mu )^4\right)}{\sqrt{2 \pi } \sigma ^7} = 0 ,\] which produces the following crossovers:

\(\left\{a_1,a_2,a_3,a_4\right\}= \)%\end{dmath*}
%\small

\begin{equation*}
%\begin{dmath*}
\left\{\mu -\sqrt{\frac{1}{2} \left(5+\sqrt{17}\right)} \sigma ,\mu -\sqrt{\frac{1}{2} \left(5-\sqrt{17}\right)} \sigma ,\\
\mu +\sqrt{\frac{1}{2} \left(5-\sqrt{17}\right)} \sigma ,
\mu +\sqrt{\frac{1}{2} \left(5+\sqrt{17}\right)} \sigma \right\}
\end{equation*}

%\normalsize
%\normalfont
In  figure 3, the crossovers for the intervals are numerically \(\{-2.13 \sigma, -.66 \sigma, .66 \sigma, 2.13 \sigma \}.\)

\ \
As to a symmetric power law(as we will see further down), the Student T Distribution with scale $s$ and tail exponent $\alpha$:

\[p (x) \equiv \frac{\left(\frac{\alpha }{\alpha +\frac{x^2}{s^2}}\right)^{\frac{\alpha +1}{2}}}{\sqrt{\alpha } s B\left(\frac{\alpha }{2},\frac{1}{2}\right)}\]

\begin{multline*}
\left\{a_1,a_2,a_3,a_4\right\}=\\
-\frac{\sqrt{\frac{5 \alpha -\sqrt{(\alpha +1) (17 \alpha +1)}+1}{\alpha -1}} s}{\sqrt{2}},
\frac{\sqrt{\frac{5 \alpha -\sqrt{(\alpha +1) (17 \alpha +1)}+1}{\alpha -1}} s}{\sqrt{2}},\\
-\frac{\sqrt{\frac{5 \alpha +\sqrt{(\alpha +1) (17 \alpha +1)}+1}{\alpha -1}} s}{\sqrt{2}},
\frac{\sqrt{\frac{5 \alpha +\sqrt{(\alpha +1) (17 \alpha +1)}+1}{\alpha -1}} s}{\sqrt{2}}
\end{multline*}

When the Student is "cubic", that is, $\alpha=3$:

\begin{equation*}
\left\{a_1,a_2,a_3,a_4\right\}=
\end{equation*}

\begin{equation*}
\left\{-\sqrt{4-\sqrt{13}} s,-\sqrt{4+\sqrt{13}} s,\\
\sqrt{4-\sqrt{13}} s,\sqrt{4+\sqrt{13}} s\right\}
\end{equation*}

We can verify that when $\alpha \to \infty $, the crossovers become those of a Gaussian. For instance, for $a_1$:
\[\lim_{\alpha \to \infty } \, -\frac{\sqrt{\frac{5 \alpha -\sqrt{(\alpha +1) (17 \alpha +1)}+1}{\alpha -1}} s}{\sqrt{2}}=-\sqrt{\frac{1}{2} \left(5-\sqrt{17}\right)} s\]

2- For some one-tailed distributions that have a "bell shape" of convex-concave-convex shape, under some conditions, the same 4 crossover points
hold. The Lognormal is a special case.

\[\left\{a_1,a_2,a_3,a_4\right\}=\]

\begin{equation*}
%\left\{
e^{\frac{1}{2} \left(2 \mu -\sqrt{2} \sqrt{5 \sigma ^2-\sqrt{17} \sigma ^2}\right)},  e^{\frac{1}{2} \left(2\mu -\sqrt{2} \sqrt{\sqrt{17} \sigma ^2+5 \sigma ^2}\right)},   
\end{equation*}

\begin{equation*}
e^{\frac{1}{2} \left(2 \mu +\sqrt{2} \sqrt{5 \sigma ^2-\sqrt{17} \sigma ^2}\right)}, e^{\frac{1}{2} \left(2 \mu +\sqrt{2} \sqrt{\sqrt{17} \sigma ^2+5 \sigma ^2}\right)} 
%\right\}
\end{equation*}

\end{document}